# Binary systems with an RR Lyrae component – progress in 2016

J. LIŠKA [1,2,3], M. SKARKA [4], Á. SÓDOR [4], ZS. BOGNÁR [4]

(1) Department of Theoretical Physics and Astrophysics, Faculty of Science, Masaryk University, Kotlářská 2, CZ-611 37 Brno, Czech Republic, jiriliska@post.cz

(2) Department of Physics, Technical University of Liberec, Studentská 2, CZ-461 17 Liberec, Czech Republic

(3) CEITEC BUT, Purkyňova 656/123, CZ-612 00 Brno, Czech Republic

(4) Konkoly Observatory, Research Centre for Astronomy and Earth Sciences, Hungarian Academy of Sciences, Konkoly Thege Miklós út. 15-17, H-1121 Budapest, Hungary, marek.skarka@csfk.mta.hu, sodor@konkoly.hu, bognar@konkoly.hu

**Abstract:** In this contribution, we summarize the progress made in the investigation of binary candidates with an RR Lyrae component in 2016. We also discuss the actual status of the RRLyrBinCan database.

**Abstrakt:** V tomto příspěvku shrnujeme pokrok, který byl proveden ve výzkumu kandidátů na dvojhvězdy s RR Lyrae složkou za rok 2016. Diskutujeme také aktuální stav RRLyrBinCan databáze.

## Introduction

Till the end of 2016 about 100 000 pulsating stars of RR Lyrae type have been identified in the Galactic field and bulge, globular clusters, SMC, and LMC. Despite that several dozen candidates for RR Lyrae stars in binary systems are known, none of them has been unambiguously confirmed. This is in clear contrast to other types of stars where binarity is very common.

More than a half of the candidates were discovered between 2014 and 2016 (e.g. Li & Qian, 2014; Hajdu et al., 2015; Liška et al., 2016a). In addition, the best candidate TU UMa was thoroughly re-analysed in Liška et al. (2016b). The reasons for the low number of known candidates were discussed e.g. in Liška et al. (2016a) and Skarka et al. (2016). In this short note we discuss candidates that appeared in literature in the second half of 2016 (were not presented in 47th Conference on Variable Stars Research), namely TU Com, KIC 2831097, RW Ari, and eclipsing systems in the LMC. We also update the status of the RRLyrBinCan database, the only up-to-date list of this kind of objects currently available.

## TU Com

TU Com is an RR Lyrae pulsator (RRab type) showing Blazhko effect. De Ponthière et al. (2016) obtained new observations with five-year time span which disproved 75-days long Blazhko period published in the General Catalogue of Variable Stars. They identified two Blazhko periods of 43.6 and 45.5 days based on maxima timings and frequency analysis. In addition, 1676-days long cycle is apparent in their and SuperWASP data which can be explained with light-travel time effect caused by orbital motion of a binary system.

## KIC 2831097

KIC 2831097 is a first-overtone RR Lyrae pulsator discovered by our team in original Kepler field (Sódor et al., 2017). We analysed stability of its pulsation period and found out that the star shows large phase variation with amplitude of 0.1 d. It can be interpreted as combination of a linear decrease of pulsation period (probably evolutionary effect) and cyclic variation explainable as light travel-time effect caused by orbital motion in a binary system. The characteristics of both processes found in KIC 2831097 are unusual. The rate of the pulsation period decrease is extreme among known RR Lyrae stars. In addition, the assumed orbital period of approximately 2 years is the shortest among the non-eclipsing RR Lyrae binary candidates. The possible companion of RR Lyrae component is a candidate for black hole due to its high mass (at least 8.4 $M_{Sun}$) and evolutionary status. In addition, the star shows numerous additional non-radial pulsation frequencies and an ~ 47-d Blazhko-like irregular light-curve modulation.

## RW Ari

RW Ari (RRc type) belongs among candidates for eclipsing systems with an RR Lyrae component for 45 years. Wiśniewski (1971) detected the eclipses in photoelectric data and found eclipsing period of 3.1754 d. After that, many authors tried to verify the eclipses, but the number of papers which confirm binarity is similar as the number of papers discarding it (see Table 1). The summary of the history of RW Ari investigation is presented e.g. in Liška et al. (2016c) or Odell & Sreedhar (2016). The last mentioned paper is also the last study focusing on binarity of RW Ari. They obtained and analysed new and archival photometric data, but they found normal brightness variation (no eclipses were detected) which is accompanied by significant changes in pulsation period. They noted at least three abrupt changes in period (one of them in 2012). Variation in radial velocity (RV) curve constructed from their data looks also normally. Probably nobody can exclude the strange features in the light curve observed by Wiśniewski (1971), but recently RW Ari behaves as a normal RRc star and not as an eclipsing binary.

Table 1. References confirming (YES) or disproving (NO) binarity of RW Ari.

| YES | NO |
|---|---|
| Wiśniewski (1971) – photometry | |
| Abt & Wiśniewski (1972) – RV | |
| Woodward (1972) – literature photometry | Penston (1972) – photometry |
| Sidorov (1978) – literature photometry | Edwards (1978) – photometry |
| | Goranskij & Shugarov (1979) – photometry |
| Dahm (1992) – literature photometry | |
| | Jeffery et al. (2007) – RV |
| | Liška et al. (2016c) – photometry |
| | Odell & Sreedhar (2016) – photometry, RV |
| **5:6** | |

## Eclipsing RR Lyrae systems in LMC

Soszyński et al. (2016) presented five eclipsing candidates in LMC and OGLE IV database. Four of them are known from their former research (Soszyński et al., 2003, 2009), only OGLE-LMC-RRLYR-30844 is newly identified system with eclipsing period of approximately 1.48 d. All of the LMC eclipsing systems have relatively short orbital periods (from 1.48 to 16.23 d) which almost exclude the possibility that a classical RR Lyrae pulsator is present in the system. Binary evolution pulsators (Pietrzyński et al., 2012) or optical blends consisting of RR Lyr star and eclipsing binary are the best explanations for observed variations.

## RRLyrBinCan database – the updates

A new database containing candidates for RR Lyrae stars in binary systems, RRLyrBinCan database, was introduced in our former papers (Liška et al., 2016a; Liška & Skarka, 2016). The list contains candidates with their basic parameters and references. CDS version from May 2016 contains 64 candidates. An updated version of the database that is available online at http://rrlyrbincan.physics.muni.cz contains 109 records for 78 stars (November 12, 2016). Sixteen of the listed stars show the Blazhko effect and six other stars are candidate Blazhko stars. Catalogized objects are mostly of RRab type (73) and the rest of them are of RRc type (5). A new utility implemented into the webpage showed us that the database is visited from the whole world.

**Conclusions**

We present a short summary of the progress in research focusing on detection and confirmation of binary systems among RR Lyrae stars in 2016. The topic is still very actual and interesting. Our team contributed with the identification and analysis of several candidates and managing the RRLyrBinCan database.

**Acknowledgement**

This research has made use of NASA's Astrophysics Data System and of the International Variable Star Index (VSX) database, operated at AAVSO, Cambridge, Massachusetts, USA. The financial support of NKFIH K-115709 and K-113117 are acknowledged. MS acknowledges the support of the postdoctoral fellowship programme of the Hungarian Academy of Sciences at the Konkoly Observatory as a host institution. ÁS was supported by the János Bolyai Research Scholarship of the Hungarian Academy of Sciences. This research was carried out under the project CEITEC 2020 (LQ1601) with financial support from the Ministry of Education, Youth and Sports of the Czech Republic under the National Sustainability Programme II. This work was supported by the Brno Observatory and Planetarium.